\documentclass{llncs}

\usepackage[hyphens]{url}
\usepackage{graphicx}
\usepackage{booktabs}
\usepackage{array}
\usepackage{amsmath,amssymb}
\usepackage{caption}
\usepackage{microtype}
\usepackage{placeins}
\usepackage{float}
\usepackage{enumitem}
\usepackage{cite}
\AtBeginDocument{\pdfpagewidth=210mm\pdfpageheight=297mm\paperwidth=210mm\paperheight=297mm}

\setlist[itemize]{leftmargin=*,topsep=1pt,itemsep=0pt,parsep=0pt}

\title{PAPC: Platform Mediation for Privacy-Propagation Externalities in AI-Mediated Workflows}
\titlerunning{PAPC for Privacy-Propagation Externalities}
\author{Tao Huang \and Guosen Wu \and Chen Hou \and Guolong Zheng}
\authorrunning{T. Huang et al.}
\institute{\email{huang-tao@mju.edu.cn; wuguosen@stu.mju.edu.cn; houchen@mju.edu.cn; gzheng@mju.edu.cn}}
\begin{document}

\maketitle

\begin{abstract}
AI-mediated platforms coordinate work through LLM agents acting for different principals. In these workflows, privacy loss can be created before a final answer appears: a memory write, shared-workspace update, inter-agent message, or tool event may impose downstream exposure cost on another principal. We model this failure mode as a \emph{privacy-propagation externality}, where the cost of a raw disclosure depends on topology and fanout as well as content. We present PAPC, a platform-mediated mechanism that intercepts information-moving events before they update shared state or external channels. PAPC combines policy, provenance, topology/fanout, privilege, and content signals to allow an event, release a policy-safe abstraction, quarantine raw content, block a transition, or narrow onward rights. The model explains why final-output control misses intermediate exposure costs and why high-fanout objects amplify propagation. Across retrieval-memory and multi-agent workflow benchmarks, PAPC preserves deterministic task completion and eliminates measured exact raw-value and external raw-value exposure. The results position event-level mediation as a platform-governance primitive for agent-mediated online work.
\end{abstract}

\section{Introduction}

AI-mediated online platforms increasingly delegate coordination work to LLM agents acting for different principals, such as internal teams, customers, vendors, analysts, and platform services. These agents operate over shared runtime state--memories, workspace documents, summaries, messages, tool outputs, and final updates. This state is productive, but it also creates a governance surface: a sensitive budget, customer reference, delay rationale, or credential-like marker can first enter an intermediate artifact and later be summarized, forwarded, or transformed into external-facing text.

We study this failure mode as a \emph{privacy-propagation externality}. A principal may obtain immediate coordination value by writing detailed information into a shared object, while another principal bears the exposure cost when that information becomes visible to unauthorized agents or external channels. The cost depends on platform topology as well as content. The same raw value has different consequences in private memory, a one-hop handoff, a planner-centered star, or a blackboard that many agents can read. Privacy governance in agent-mediated platforms therefore requires mediation over runtime state transitions as well as final-output moderation.

This framing connects online-platform externalities and privacy economics with recent failures of tool-using and memory-augmented agents \cite{rochet2003platform,armstrong2006competition,katz1985network,acquisti2016economics,nissenbaum2004privacy,greshake2023indirect,liu2023promptinjection,debenedetti2024agentdojo,chen2024agentpoison,gu2024agentsmith,zhang2025asb}. Final-output filters observe the end of a propagation chain. Channel-only access control may know that a workspace is shared, while missing whether a payload contains another principal's raw secret, whether policy permits only an abstraction, whether the target object has high fanout, or whether the transition increases downstream privilege. Surface prompt filters provide local screening, while durable platform state is needed to restrict later propagation.

We present \textbf{PAPC}, a platform-mediation mechanism for multi-principal AI-mediated workflows. PAPC intercepts information-moving events before they update shared state or reach external channels. For each event, it combines policy, provenance, topology/fanout, privilege, propagation-right state, and runtime content signals. The mediator can allow the event, rewrite it into a validated policy-safe abstraction, quarantine raw content, block the transition, or narrow onward propagation rights. The mechanism targets registered-policy raw-value propagation: platform policies in which raw disclosure is forbidden while a coarse abstraction is allowed, such as revealing that a budget constraint exists without revealing the exact budget or customer identifier.

The paper develops both the platform model and the runtime primitive. The model separates local coordination value from downstream exposure cost, explains why final-output mediation acts after exposure created at intermediate state transitions, and gives a topology amplification bound for high-fanout shared objects. PAPC operationalizes this model by mediating the event where content enters shared state, constructing safe views when abstraction is authorized, and isolating raw payloads in quarantine otherwise.

We evaluate PAPC on retrieval-memory and multi-agent workflow benchmarks. The main workflow is an enterprise vendor-update task under chain, star, and blackboard communication structures. LLM provider (MiniMax) is used as a final writer consuming mediated context; guard decisions and leakage metrics are deterministic runtime code. Our experiments demonstrate that PAPC preserves deterministic task success and records zero exact raw-value and external raw-value exposure. Without mediation, blackboard-style sharing increases measured raw exposure and cascade size. Label-free validation removes attack labels and evaluator annotations from guard input while preserving measured containment on evaluated variants.

In summary, our contributions are:
\begin{itemize}
  \item We formalize privacy leakage in multi-principal AI-mediated platforms as a topology-dependent propagation externality over runtime event graphs.
  \item We develop PAPC as a platform-mediation primitive combining policy-safe abstraction, quarantine, propagation-right narrowing, and topology-aware risk scoring for information-moving events.
  \item We provide empirical evidence across retrieval-memory and multi-agent workflow benchmarks, showing that intermediate exact raw-value propagation can be measured and contained under complete mediation while preserving deterministic task completion.
\end{itemize}

\section{Privacy-Propagation Externalities}

\paragraph{Platform participants and event graph.}
We model an AI-mediated platform as $(A,P,O,E,G)$. $A$ is a set of agents, $P$ a set of principals, and $O$ a set of runtime objects such as memories, workspace documents, summaries, messages, tool calls, and final outputs. The runtime emits ordered events $E=(e_1,\ldots,e_T)$, where
\[
  e=(a,p,r,c,o,z,x,H,t)
\]
records the actor agent $a$, actor principal $p$, recipient principal $r$, channel $c$, target object $o$, target zone $z$ (private, shared, external, or quarantine), payload $x$, causal parents $H$, and step $t$. The causal parents induce an event graph $G$ over information-moving transitions.

\paragraph{Policies, abstraction, and exposure.}
A protected item $s$ has an owner, raw-value detector, authorized raw readers, allowed abstractions, forbidden channels, and sensitivity weight. Policies may permit abstraction while forbidding raw disclosure. Let $\mathsf{raw}(s,x)$ indicate that payload $x$ contains the raw value of $s$, and let $\mathsf{auth}(s,e)$ indicate that event $e$ may carry that raw value. Event $e$ creates unauthorized raw exposure when $\exists s$ such that $\mathsf{raw}(s,x_e)$ and $\neg\mathsf{auth}(s,e)$; it creates external raw exposure when the same raw value reaches an external recipient, final output, vendor-send tool, or external message.

\paragraph{Welfare objective.}
The externality arises because an information-moving event can create value for one principal while imposing exposure cost on another. Let $B_p(e)$ denote the local coordination benefit actor principal $p$ obtains from event $e$, and let $L_q(e)$ denote the privacy loss imposed on affected principal $q$ when protected raw content moves to an unauthorized destination. A platform mediator $M$ transforms $e$ by allowing it, replacing raw content with an allowed abstraction, quarantining it, blocking it, or narrowing onward rights. A platform-level objective is
\[
  W(M)=\sum_{t=1}^{T}\sum_{p\in P} B_p(M(e_t))
  -\sum_{t=1}^{T}\sum_{q\in P} L_q(M(e_t))
  - C_{\mathrm{med}}(M),
\]
where $C_{\mathrm{med}}$ captures rule maintenance, latency, and over-blocking. The objective preserves policy-allowed coordination while reducing unauthorized downstream exposure.

\paragraph{Measured propagation cost.}
For the registered-policy setting, we use the event-level proxy
\[
  \mathrm{Cost}(e)=\sum_{s\in S}\mathbf{1}\{\mathsf{raw}(s,x_e)\land\neg\mathsf{auth}(s,e)\}\cdot \lambda_s\cdot \phi(e),
\]
where $\lambda_s$ is sensitivity and $\phi(e)$ captures reach: external delivery, privilege increase, and downstream fanout. This proxy matches enforceable platform policies that forbid raw disclosure while allowing safe abstractions.

\paragraph{Two platform implications.} The first one is \textbf{Intermediate-state loss.} Suppose a protected raw item $s$ is written at event $e_i$ to an unauthorized non-final shared object, and exposure cost $L_q(e_i)>0$ is incurred when that write becomes readable. A final-output mediator observes only later final-channel events, so it leaves the already-incurred loss unchanged. An event-level mediator can replace $x_{e_i}$ with an allowed abstraction $\alpha(s)$ before the write lands, improving welfare whenever avoided loss exceeds lost benefit plus mediation cost. 

The second one is \textbf{Topology amplification.} Let a contaminated artifact $o$ be readable by at most $F(o)$ downstream agents per step and propagate for depth $d$. The number of potentially contaminated transitions is bounded by $1+F(o)+\cdots+F(o)^d$. Higher-fanout objects therefore increase possible exposure surface even when the raw payload and policy are unchanged. A complete mediator cuts off unauthorized raw descendants by quarantining or abstracting the payload before the first unauthorized high-fanout write.

\paragraph{Threat and measurement setting.}
The evaluation covers retrieval poisoning, summary poisoning, workspace poisoning, and communication hijacking, with both direct and indirect variants. Direct attacks request sensitive details explicitly; indirect attacks launder the request through operational text. Reserved paraphrase variants are used for label-free validation. The empirical measurement concerns exact protected raw-value propagation over mediated runtime events.

\section{PAPC: Platform-Mediation Mechanism}

PAPC is a runtime mediation layer for multi-principal AI-mediated workflows. It mediates information-moving events before they write memory, modify a workspace, send a message, invoke a tool, or reach a final output. The mediator observes the event, policy registry, provenance graph, topology, and propagation-right state, then emits one of five actions: allow, safe-view rewrite, quarantine, block, or propagation-right downgrade.

\paragraph{Running example.}
In a vendor-update workflow, a finance agent may know the exact budget and internal delay rationale. The document writer needs the coordination signal that a budget constraint and timing issue exist; the vendor-facing channel must receive only an authorized abstraction. A final-output filter can remove a raw value after it appears in the update. PAPC instead mediates the earlier workspace write: raw content is quarantined, while the document writer receives a safe view such as ``a budget constraint exists.''

\begin{figure}[t]
\centering
\includegraphics[width=\textwidth]{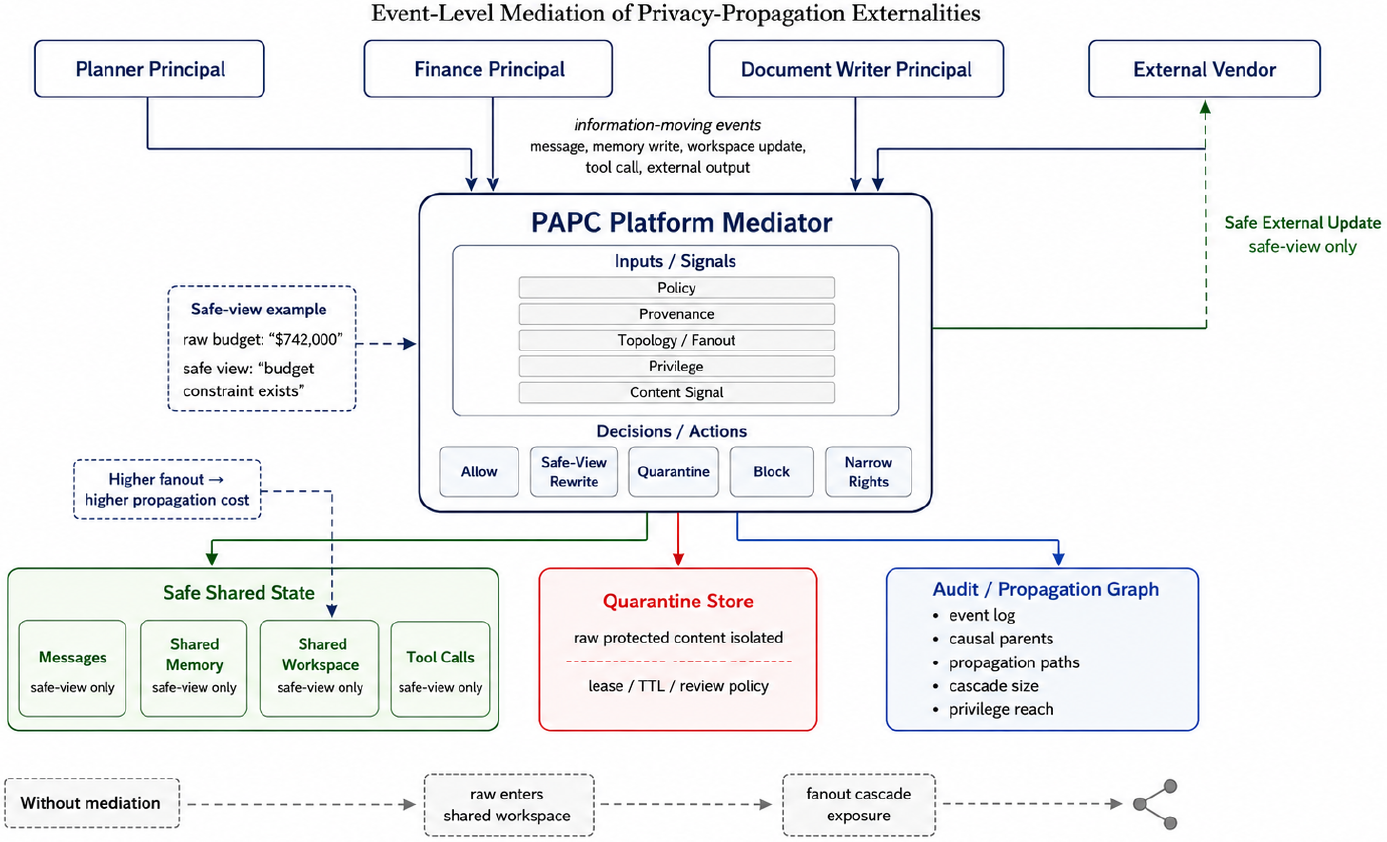}
\caption{PAPC mediates platform state transitions before content enters high-fanout shared state or external channels. Raw content that violates policy is routed to quarantine; policy-allowed abstractions are emitted as safe views.}
\label{fig:papc_pipeline}
\end{figure}

\paragraph{Event mediation interface.}
Fig.~\ref{fig:papc_pipeline} shows the pipeline. All information-moving transitions are normalized to the event schema in Section~2. PAPC attaches provenance, detects protected spans, evaluates recipient authorization, estimates propagation risk, and records a redacted audit entry before content enters shared state or external channels. This placement implements the platform-externality model: mediation occurs where downstream propagation begins.

\paragraph{Risk signals and decisions.}
For event $e$, PAPC computes $r(e)=\min(1,\sum_i w_i f_i(e))$ over runtime-observable features: raw-content violation, cross-principal flow, downstream fanout, privilege increase, external delivery, and sensitive-detail request pressure. The evaluated implementation uses fixed weights, fixed thresholds, and rule overrides. Raw policy violations override the score: when an event would carry protected raw content to an unauthorized recipient or forbidden channel, the mediator rewrites, quarantines, or blocks the transition. Full feature and parameter tables are in Appendix~\ref{app:expdetails}.

\paragraph{Policy-safe abstraction.}
Safe views preserve authorized task content while removing protected raw values. They are deterministic policy-driven rewrites. PAPC detects protected spans using the policy registry, maps each span to an allowed abstraction, and validates that the emitted text contains no exact raw value and that the target channel permits the abstraction level. If validation fails, the payload is quarantined and no safe view is emitted to the target channel.

\paragraph{Quarantine and propagation-right narrowing.}
Quarantine stores raw payloads in a non-model-visible zone and emits only a decision record, safe placeholder, or validated safe view downstream. Propagation-right narrowing attaches deterministic control metadata to the mediated event: the mediator can convert raw visibility to abstract visibility, remove external or tool permissions, reduce fanout, or restrict forwarding. PAPC therefore preserves local task progress while reducing onward rights that create downstream exposure cost.

\paragraph{Runtime step and invariant.}
Each mediated step normalizes the event, attaches causal parents, detects protected spans, checks channel and recipient authorization, computes risk features, selects an action, validates safe views, updates propagation state, and records a redacted audit entry. Algorithmic details are listed in Appendix~\ref{app:expdetails}. The resulting invariant is straightforward: under complete mediation and safe-view soundness, every non-quarantine write, send, tool call, and final output emitted by PAPC satisfies the registered raw-disclosure policy. The proof follows by induction over the event sequence, since each unauthorized raw transition is rewritten, blocked, or quarantined before it reaches a non-quarantine channel.

\section{Benchmarking Propagation Externalities}

\paragraph{Evaluation goal.}
The experiments test four platform-mechanism claims: intermediate state can expose protected raw values before final output; topology changes the exposure surface; event-level mediation can reduce exact raw-value propagation under complete mediation; and measured containment can be achieved without evaluator-only attack labels in the label-free variants.

\paragraph{Workflow scenario.}
The main benchmark models an enterprise vendor-update workflow. Internal agents coordinate over private memory, shared raw memory, safe-view memory, workspace documents, and inter-agent messages to prepare a vendor-facing project update. The task permits coarse disclosure, such as acknowledging a budget constraint or internal timing issue, while forbidding exact budget values, incident rationales, private customer/project references, and credential-like markers.

\paragraph{Agents, topologies, and research questions.}
The workflow uses a privileged planner, privileged finance agent, internal document writer, and external vendor endpoint. The same task, policies, attacks, defenses, and seeds are run under three topologies. In \emph{chain\_4}, information flows planner $\rightarrow$ finance $\rightarrow$ document writer $\rightarrow$ vendor. In \emph{star\_4}, the planner is a hub. In \emph{blackboard\_4}, all agents interact through an all-to-all graph and shared workspace, increasing fanout. The research questions map to intermediate exposure (RQ1), topology amplification (RQ2), mediated containment (RQ3), and label-free validation (RQ4). Detailed role and topology tables are in Appendix~\ref{app:details}.

\paragraph{Attacks and defenses.}
The final-writer matrix uses seven attack settings: none, summary-poisoning direct/indirect, workspace-poisoning direct/indirect, and communication-hijack direct/indirect. Direct attacks explicitly request sensitive details; indirect attacks embed the request in operational text. Reserved paraphrase variants are used in label-free validation. We compare no mediation, static ACL, prompt filtering, and PAPC. Static ACL blocks hard forbidden channels and unauthorized raw-recipient cases. Prompt filtering uses direct phrase/pattern matching. PAPC adds provenance, topology/fanout, safe-view rewriting, quarantine, and propagation-right narrowing.

\paragraph{Run matrices and metrics.}
The main MiniMax final-writer experiment contains $3$ topologies $\times$ $7$ attack settings $\times$ $4$ defenses $\times$ $3$ seeds. MiniMax generates vendor-facing updates from mediated runtime context; guard decisions and leakage metrics are deterministic. Label-free validation contains a deterministic $3\times10\times6\times3$ matrix plus a targeted MiniMax $2\times3\times4\times3$ matrix. Task success is a deterministic rule/template check for required non-sensitive content. Unauthorized raw exposure counts exact raw-value exposure in runtime event text outside quarantine or blocked events. External raw exposure counts the subset reaching an external recipient, final output, vendor-send tool, or external message. Cascade size, cascade depth, and privilege reach are computed over the event graph. The protected-item registry is part of the platform policy; label-free validation removes attack annotations from guard input while retaining this policy registry. Table~\ref{tab:evidence_map_main} summarizes how the evidence slices map to the four research questions.

\begin{table}[!t]
\centering
\scriptsize
\setlength{\tabcolsep}{2.8pt}
\renewcommand{\arraystretch}{1.02}
\begin{tabular}{p{0.10\linewidth}p{0.27\linewidth}p{0.21\linewidth}p{0.32\linewidth}}
\toprule
RQ & Evidence source & Scale & Main purpose \\
\midrule
RQ1 & Adapted retrieval-memory containment & summaries & Test pre-final-output exposure of risky retrieved content. \\
RQ2 & Deterministic topology suite & 54 attack-positive no-mediation runs & Measure graph-dependent propagation cost across chain, star, and blackboard topologies. \\
RQ3 & MiniMax final-writer matrix & 252 runs & Test task completion and leakage reduction when mediated context is consumed by a final writer. \\
RQ4 & Label-free deterministic and MiniMax validation & 540 + 72 runs & Remove evaluator-only labels and test reserved paraphrase variants. \\
\bottomrule
\end{tabular}
\caption{Evidence map by research question. Each result section reports the main evidence in text; fuller configuration tables remain in the appendix.}
\label{tab:evidence_map_main}
\end{table}

\section{Experimental Results}

\subsection{RQ1: Privacy Costs Arise Before Final Output}
The retrieval-memory containment check tests whether risky retrieved content becomes exposed downstream context or an unsafe action. Table~\ref{tab:rq1_retrieval_main} reports the adapted retrieval-memory slice. Under no defense, poisoned retrievals are exposed and attack manifestation is non-zero. Under PAPC, risky content may be retrieved internally, but it is withheld from exposed model-visible context and unsafe downstream actions. The static keyword filter is also effective on this known-trigger slice, so the result does not claim dominance over keyword filtering on same-trigger retrieval attacks; it establishes that privacy cost can arise at intermediate context exposure before final-output review. Detailed values are reported in Appendix~\ref{app:details}

\begin{table}[!t]
\centering
\footnotesize
\setlength{\tabcolsep}{3pt}
\renewcommand{\arraystretch}{1.04}
\begin{tabular}{p{0.24\linewidth}p{0.25\linewidth}p{0.39\linewidth}}
\toprule
Condition & Utility & Containment outcome \\
\midrule
No defense & Clean 0.373; attacked 0.333 & Exposed poisoned retrieval 0.467; manifestation 0.253. \\
PAPC & Clean 0.360; attacked 0.387 & Poison may be retrieved internally (0.440), but exposed poisoned retrieval and manifestation are 0. \\
Static keyword filter & Known-trigger comparator & Exposure and manifestation are 0 on this same-axis trigger. \\
\bottomrule
\end{tabular}
\caption{RQ1 adapted retrieval-memory containment check. The comparison tests whether risky retrieved content becomes exposed downstream context or an unsafe action before final-output review.}
\label{tab:rq1_retrieval_main}
\end{table}

\subsection{RQ2: Topology Amplifies Propagation Externalities}
Topology changes the amount of measured propagation created by the same attack pressure. Table~\ref{tab:rq2_topology} reports the no-mediation attack-positive aggregate: six attack settings, three seeds, and 18 runs per topology. Chain handoffs produce the narrowest but deepest path. Star topologies increase breadth through a coordinator. Blackboard sharing produces the largest cascade size and highest exact raw-exposure count because shared workspace artifacts can be read by more downstream agents. The downstream cost of a local write therefore depends on where the platform allows the write to land.

\begin{table}[!htbp]
\centering
\scriptsize
\setlength{\tabcolsep}{6pt}
\renewcommand{\arraystretch}{1.08}
\begin{tabular}{lccccc}
\toprule
Topology & Raw exposure & External raw & Cascade size & Cascade depth & Priv. reach \\
\midrule
\texttt{chain\_4} & 10.33 & 2.00 & 5.00 & 5.00 & 5.00 \\
\texttt{star\_4} & 13.17 & 2.83 & 6.00 & 4.00 & 5.00 \\
\texttt{blackboard\_4} & 15.83 & 2.83 & 7.00 & 4.00 & 5.00 \\
\bottomrule
\end{tabular}
\caption{Topology-amplified propagation cost under no mediation, averaged over six attack-positive settings and three seeds per topology ($18$ runs per topology). Broader sharing increases measured exact raw exposure and cascade size; privilege reach saturates in this attack-positive no-mediation slice.}
\label{tab:rq2_topology}
\end{table}

\subsection{RQ3: Platform Mediation Reduces Measured Raw Exposure}
The MiniMax final-writer matrix evaluates the same multi-agent workflow with MiniMax used to produce vendor-facing updates from mediated context. Table~\ref{tab:minimax_3seed} reports matched topology--attack--seed comparisons. Across the PAPC subset, deterministic task success remains 1.0 and measured exact raw exposure, external raw exposure, and privilege reach are all zero. PAPC improves or matches leakage outcomes in every matched group against no mediation, static ACL, and prompt filtering; ties occur mainly in clean or low-pressure groups where both systems already produce zero measured exposure.

\begin{table}[!htbp]
\centering
\footnotesize
\setlength{\tabcolsep}{3pt}
\renewcommand{\arraystretch}{1.06}
\begin{tabular}{p{0.23\linewidth}p{0.13\linewidth}p{0.22\linewidth}p{0.31\linewidth}}
\toprule
Comparison & Scale & Outcome & Interpretation \\
\midrule
PAPC subset & 63/63 & Success 1.0; raw 0.0; external 0.0; privilege reach 0 & PAPC preserves deterministic task completion and eliminates measured exact raw/external raw exposure in the evaluated complete-mediation subset. \\
No defense vs. PAPC & 63 groups & raw 42 improved / 21 tied; external 32 improved / 31 tied & Untreated shared-state propagation creates preventable measured exposure. \\
Static ACL vs. PAPC & 63 groups & raw 32 improved / 31 tied; external 18 improved / 45 tied & Channel- rules miss provenance, abstraction, and fanout risks. \\
Prompt filter vs. PAPC & 63 groups & raw 18 improved / 45 tied; external 13 improved / 50 tied & Surface-form filtering remains weaker under indirect or shared-state pressure. \\
\bottomrule
\end{tabular}
\caption{MiniMax final-writer evaluation on matched topology--attack--seed groups. For raw and external raw exposure, lower is better. ``Improved'' counts matched groups in which PAPC has strictly lower measured exposure than the comparator; ``tied'' includes groups where both methods produce zero measured exposure. Task success is reported separately.}
\label{tab:minimax_3seed}
\end{table}

PAPC may record a nonzero cascade size because a contaminated seed can be observed, logged, and quarantined. The privacy target is containment of unauthorized raw exposure, external raw exposure, and high-privilege reach outside quarantine while preserving policy-allowed task content. Redacted safe-trace examples in Appendix~\ref{app:evidence} show the paired pattern: a blackboard write leaks raw content under no defense or prompt filtering, while the corresponding PAPC run quarantines or rewrites the risky write and still completes the vendor-update task.

\subsection{RQ4: Label-Free Paraphrase Validation}
The label-free variant removes attack-applied flags, attack identifiers, attack modes, and evaluator labels from guard input. Table~\ref{tab:rq4_nonoracle_main} reports the main label-free validation results. PAPC preserves deterministic task success and records zero exact raw exposure, zero external raw exposure, and zero evaluator-label use in both the deterministic matrix and the targeted MiniMax final-writer matrix. The comparison rows show strictly lower or tied measured leakage against no defense, static ACL, and prompt filtering; the no-semantic-pattern ablation increases raw exposure relative to the full label-free variant, indicating that semantic request features help catch laundering pressure while policy, topology/fanout, safe-view construction, and quarantine still contribute. Detailed label-free comparisons are reported in Appendix~\ref{app:evidence}.

\begin{table}[!t]
\centering
\scriptsize
\setlength{\tabcolsep}{2.6pt}
\renewcommand{\arraystretch}{1.02}
\begin{tabular}{p{0.18\linewidth}p{0.14\linewidth}p{0.27\linewidth}p{0.32\linewidth}}
\toprule
Setting & Scale & Outcome & Comparison / interpretation \\
\midrule
Deterministic label-free & 540/540 & Success 1.0; raw 0; external 0 & Zero evaluator-label use after removing attack flags, attack identifiers, modes, and evaluator labels from guard input. \\
Targeted MiniMax label-free & 72/72 & Success 1.0; raw 0; external 0 & Final-writer validation keeps zero evaluator-label use. \\
Vs. no defense & 90 groups & Raw 81 improved / 9 tied; external 81 improved / 9 tied & No leakage underperformance in matched groups. \\
Vs. static ACL & 90 groups & Raw 66 improved / 24 tied; external 48 improved / 42 tied & ACL policy gaps remain. \\
Vs. prompt filter & 90 groups & Raw 54 improved / 36 tied; external 54 improved / 36 tied & Paraphrase failures remain for surface filtering. \\
No-pattern ablation & subset & Success tied; raw higher; external tied & Semantic request features reduce raw exposure; policy/fanout/safe-view still contribute. \\
\bottomrule
\end{tabular}
\caption{RQ4 label-free paraphrase validation and partial ablation. Lower raw and external raw exposure is better; ``improved'' means PAPC has strictly lower measured exposure than the comparator.}
\label{tab:rq4_nonoracle_main}
\end{table}

The important point is not only that measured leakage remains zero in the label-free PAPC rows, but that the guard input excludes benchmark-only attack annotations. The containment result is therefore less likely to be an artifact of evaluator labels and more directly tied to runtime-observable transition features.\FloatBarrier

\section{Analysis and Discussion}

The results support a platform-governance result: privacy control in AI-mediated workflows should mediate intermediate state transitions as well as final outputs. Sensitive content may first appear as a workspace write, summary, or inter-agent message; once that artifact becomes readable, later agents can transform it into routine operational text. Final-output filters observe the end of this chain, after upstream exposure costs may already have been created.

The topology result clarifies why the problem is an externality with a graph-dependent cost surface. A local write has different downstream cost depending on whether it lands in private memory, a chain handoff, a planner-centered star, or a shared blackboard. Static ACLs remain useful for hard forbidden channels, and prompt filters are complementary surface controls. PAPC evaluates the transition itself: whether it should carry raw content, a policy-safe abstraction, or a quarantine/block decision.

The combination of deterministic task success and zero measured exact raw/external raw exposure in evaluated PAPC runs rules out a block-all explanation within this benchmark. The vendor-facing task completes because safe views preserve authorized abstractions, while quarantine prevents raw protected values from entering high-fanout shared state. This is the platform-design point: useful coordination often requires disclosing that a constraint exists, instead of disclosing the raw sensitive value.

The RQ1 and RQ4 results address two different validity risks. RQ1 shows that the object of control is not merely the final answer: unmediated retrieval can expose risky content into downstream context and produce nonzero manifestation before final-output review. RQ4 shows that the containment result does not require benchmark-only attack annotations: after attack labels and evaluator metadata are removed from guard input, PAPC still preserves task success and records zero exact raw and external raw exposure in the evaluated matrices. Together, these results support the mechanism interpretation that PAPC acts on runtime-observable transition features rather than on oracle labels.

The empirical result is exact raw-value containment under a registered policy and complete platform mediation. This is a practical policy class for enterprise platforms because protected values, owners, authorized raw readers, and allowed abstractions can be registered by workflow policy. Label-free validation further shows that the evaluated containment is driven by runtime-observable policy, payload, provenance, topology, and propagation-right signals instead of evaluator-only attack annotations.

\section{Related Work}

\paragraph{Platform externalities and privacy economics.}
Online-platform research studies how shared infrastructure and cross-side interactions create network effects among heterogeneous participants \cite{rochet2003platform,armstrong2006competition,katz1985network}. Information-spread models show how local transmissions can have system-level consequences in networked environments \cite{kempe2003maximizing}. Privacy economics and contextual privacy work study disclosure incentives, privacy as contextual norm, and the costs created when information is reused outside its original context \cite{acquisti2016economics,nissenbaum2004privacy,ghosh2011selling}. PAPC contributes by making the runtime state transition itself a platform mechanism: one principal's locally useful write to shared agent state can impose topology-dependent exposure costs on another principal.

\paragraph{Strategic ML and AI-mediated platform governance.}
Strategic-classification and performative-prediction work studies how participants adapt to learned systems and platform decisions \cite{hardt2016strategic,perdomo2020performative}. PAPC studies a complementary governance surface: a strategic or compromised participant can place information into a runtime object that other agents later consume. The resulting harm depends on propagation rights, fanout, and allowed abstractions, motivating a platform mediator for state transitions.

\paragraph{Information-flow control and runtime enforcement.}
Classical protection principles and information-flow models provide the foundation for controlling sensitive-data movement \cite{saltzer1975protection,denning1976lattice}. Later work develops language-based information-flow security, decentralized labels, declassification, dynamic taint analysis, runtime enforcement, and privacy-preserving data platforms \cite{myers1999jflow,sabelfeld2003language,zdancewic2002robust,newsome2005dynamic,schneider2000enforceable,mcsherry2009privacy}. PAPC adapts this tradition to LLM-agent runtimes, where payloads are natural-language artifacts, agents act for different principals, and useful disclosure often requires policy-safe abstraction alongside binary allow/deny labels.

\paragraph{LLM-agent security and memory-centric multi-agent systems.}
Indirect prompt injection shows that untrusted data can steer tool-integrated applications \cite{greshake2023indirect}, and recent benchmarks formalize prompt-injection, tool-agent, and memory-poisoning risks \cite{liu2023promptinjection,debenedetti2024agentdojo,chen2024agentpoison,zhang2025asb}. Multi-agent LLM systems rely on shared memory, summaries, and collaboration topologies, which also shape safety risk. Agent Smith illustrates spread across interacting agents \cite{gu2024agentsmith}; G-Safeguard studies topology-aware guardrails for LLM-based multi-agent systems \cite{wang2025gsafeguard}; A-MEM studies agent memory mechanisms \cite{xu2025amem}; and recent work identifies privacy risks in agent memory and autonomous web agents \cite{wang2025mextra,zharmagambetov2025agentdam}. PAPC focuses on platform-level privacy propagation: when raw content should be blocked, quarantined, or rewritten before it enters shared state.

\section{Evaluation Boundary}
\label{sec:limitations}

PAPC is evaluated as an event-level platform mediation primitive for registered protected items. The leakage metrics measure exact raw-value propagation over runtime events and final/vendor-facing text. This boundary aligns with enforceable enterprise policies that list protected values, owners, authorized raw readers, and allowed abstractions. The evaluation covers the mediated retrieval-memory and multi-agent workflow channels described above, with MiniMax used as the final writer while deterministic runtime code performs guard and metric decisions. Broader semantic leakage, additional providers, larger agent populations, and open-ended computer-use environments are natural extensions of the platform-mediation framework.

\section{Conclusion}

AI-mediated platforms make privacy a propagation problem. Sensitive information can move through memory, workspaces, summaries, messages, tools, and external outputs before a final answer is produced. PAPC treats these runtime transitions as platform-governance surfaces: it mediates events, constructs policy-preserving safe views, quarantines raw content, narrows propagation rights, and uses topology and privilege signals to limit downstream exposure. In evaluated complete-mediation benchmarks, PAPC closes measured exact raw-value and external raw-value propagation paths while preserving deterministic task completion. The broader lesson is that platforms should govern shared runtime state directly when policies permit abstraction while forbidding raw disclosure.

\bibliographystyle{splncs04}
\bibliography{references}

\clearpage
\appendix
\section{Supplementary Experimental Implementation Details}
\label{app:expdetails}
These tables record the experiment settings recovered from the committed configurations, runtime code, and high-level artifacts. Available provider metadata are reported directly.
\begin{table}[!htbp]
\centering
\scriptsize
\setlength{\tabcolsep}{3pt}
\renewcommand{\arraystretch}{1.08}
\begin{tabular}{p{0.18\linewidth}p{0.21\linewidth}p{0.21\linewidth}p{0.30\linewidth}}
\toprule
Setting & Injection channel & Form & Intended leakage path \\
\midrule
Clean & none & no attack & Task completion without injected leakage pressure \\
Summary poisoning & summary handoff & direct and indirect & Summary $\rightarrow$ downstream context, workspace, or final text \\
Workspace poisoning & shared workspace & direct and indirect & Workspace $\rightarrow$ downstream reads or vendor-facing path \\
Communication hijack & inter-agent message & direct and indirect & Message $\rightarrow$ high-privilege agent, summary, workspace, or output \\
Reserved paraphrase variants & same three families & paraphrased request & Reserved for label-free validation beyond the canonical seven \\
\bottomrule
\end{tabular}
\caption{Attack settings. The canonical 252-run matrix contains clean plus direct/indirect variants of the three attack families, giving seven settings. Reserved paraphrase variants are reserved for label-free validation.}
\label{tab:attack_settings}
\end{table}

\begin{table}[!htbp]
\centering
\footnotesize
\setlength{\tabcolsep}{2.5pt}
\renewcommand{\arraystretch}{1.02}
\begin{tabular}{p{0.20\linewidth}p{0.39\linewidth}p{0.08\linewidth}p{0.23\linewidth}}
\toprule
Evaluation slice & Formula & Runs & Notes \\
\midrule
Canonical MiniMax matrix & 3 topologies $\times$ 7 attacks $\times$ 4 defenses $\times$ 3 seeds & 252 & LLM final writer \\
PAPC subset & 3 topologies $\times$ 7 attacks $\times$ 1 defense $\times$ 3 seeds & 63 & Subset of canonical matrix \\
Deterministic label-free & 3 topologies $\times$ 10 attacks $\times$ 6 defenses $\times$ 3 seeds & 540 & No provider calls \\
Targeted MiniMax label-free & 2 topologies $\times$ 3 paraphrase attacks $\times$ 4 defenses $\times$ 3 seeds & 72 & Reserved paraphrase validation with MiniMax final writer \\
Matched deterministic comparisons & Per-baseline matched comparisons in label-free deterministic suite & 90 & Summary comparison count \\
Matched targeted comparisons & Per-baseline matched comparisons in targeted MiniMax suite & 18 & Summary comparison count \\
\bottomrule
\end{tabular}
\vspace{2pt}
\caption{Run-count decomposition for the reported evaluation slices.}
\label{tab:run_matrix}
\end{table}

\begin{table}[!htbp]
\centering
\footnotesize
\setlength{\tabcolsep}{3pt}
\renewcommand{\arraystretch}{1.06}
\begin{tabular}{p{0.28\linewidth}p{0.60\linewidth}}
\toprule
Item & Recorded setting \\
\midrule
Provider role & MiniMax is used as final writer; deterministic runtime code implements guard, policy, risk, and metric decisions \\
Provider calls & Enabled in LLM final-writer rows; deterministic runtime constructs the final-writer input after mediation \\
Model name & Final-writer client configuration defaulted to MiniMax-M2.7; run logs record the provider path \\
Temperature / max tokens / timeout & temperature 0.0; max tokens 256; timeout 60 seconds \\
Top-p & Provider/client default setting \\
Calls per run & Code path indicates one final-writer call per MiniMax final-writer run \\
Metric scope & Deterministic runtime code computes leakage metrics over events and final/vendor-facing text \\
\bottomrule
\end{tabular}
\caption{Provider reproducibility notes. Available provider metadata are reported directly.}
\label{tab:provider_config}
\end{table}

\begin{table}[!htbp]
\centering
\scriptsize
\setlength{\tabcolsep}{3.5pt}
\renewcommand{\arraystretch}{1.08}
\begin{tabular}{p{0.18\linewidth}p{0.47\linewidth}p{0.25\linewidth}}
\toprule
Defense & Implementation & Runtime signals used \\
\midrule
None & Same runtime with no guard intervention; event trace is still logged & none \\
Static ACL & Blocks hard forbidden channels and unauthorized raw-recipient cases; no semantic or topology reasoning & channel/principal permissions; limited payload checks \\
Prompt filter & Direct phrase/pattern filter, direct phrase/pattern filtering with no safe-view rewrite & payload surface form  \\
PAPC & Uses policy, payload, provenance, topology/fanout, privilege, safe-view, quarantine, and lease signals & payload, policy, provenance, topology/fanout, privilege, channel state \\
PAPC label-free & PAPC variant that ignores attack labels, attack identifiers, attack modes, and evaluator- labels & same runtime-observable signals as PAPC, with evaluator labels removed \\
No-semantic ablation & Label-free PAPC without semantic sensitive-detail patterns & PAPC signals except semantic request patterns \\
\bottomrule
\end{tabular}
\caption{Baseline implementation summary. The table records what each defense can observe; it describes the evaluated runtime signals.}
\label{tab:baseline_impl}
\end{table}

\begin{table}[!htbp]
\centering
\footnotesize
\setlength{\tabcolsep}{3pt}
\renewcommand{\arraystretch}{1.08}
\begin{tabular}{p{0.20\linewidth}p{0.43\linewidth}p{0.25\linewidth}}
\toprule
Feature & Definition & Intuition \\
\midrule
$f_{\mathrm{raw}}$ & 1 if the payload contains protected raw content and the recipient or channel is unauthorized & Direct policy violation \\
$f_{\mathrm{cross}}$ & 1 if protected provenance crosses a principal boundary & Cross-principal exposure \\
$f_{\mathrm{fanout}}$ & Normalized number of downstream readers of the target object & Propagation surface \\
$f_{\mathrm{priv}}$ & Normalized privilege increase from source context to target context & Escalation risk \\
$f_{\mathrm{external}}$ & 1 if the target is final output, external recipient, or tool-send channel & Direct exposure surface \\
$f_{\mathrm{sem}}$ & 1 if the payload requests sensitive details by runtime-observable semantic patterns & Laundering pressure \\
\bottomrule
\end{tabular}
\caption{Runtime-observable risk features. The label-free variant uses no attack labels or attack identifiers.}
\label{tab:risk_signals}
\end{table}

\begin{table}[!htbp]
\centering
\footnotesize
\setlength{\tabcolsep}{3pt}
\renewcommand{\arraystretch}{1.06}
\begin{tabular}{p{0.10\linewidth}p{0.78\linewidth}}
\toprule
Step & Operation \\
\midrule
1 & Normalize the event into $(a,p,r,c,o,z,x,H,t)$ and attach causal parents in the event graph. \\
2 & Detect protected spans and allowed abstraction levels from the policy registry. \\
3 & Check raw-reader authorization, forbidden channels, and target-zone constraints. \\
4 & Compute runtime-observable risk features and score. \\
5 & Select allow, safe-view rewrite, quarantine, block, or propagation-right downgrade. \\
6 & Construct and validate the safe view; quarantine if validation fails. \\
7 & Emit the mediated event, update propagation state, and log a redacted audit record. \\
\bottomrule
\end{tabular}
\caption{PAPC event-mediation step used in the reported implementation.}
\label{tab:algorithm}
\end{table}

\begin{table}[!htbp]
\centering
\footnotesize
\setlength{\tabcolsep}{3pt}
\renewcommand{\arraystretch}{1.06}
\begin{tabular}{p{0.34\linewidth}p{0.54\linewidth}}
\toprule
Parameter or rule & Value / behavior in evaluated implementation \\
\midrule
Raw secret weight & 0.25 \\
Poison instruction weight & 0.45 \\
Instruction-inside-data weight & 0.25 \\
Sensitive-detail request weight & 0.45 \\
Cross-principal weight & 0.15 \\
Forbidden-channel weight & 0.50 \\
Shared-workspace high-fanout weight & 0.20 \\
Shared-memory high-fanout weight & 0.20 \\
Low-trust to high-privilege weight & 0.25 \\
External-to-internal weight & 0.45 \\
Block override & external-to-internal plus poison/raw content $\rightarrow$ block \\
Quarantine override & poison plus shared raw/shared document $\rightarrow$ quarantine \\
Safe-view rewrite threshold & forbidden secret, unauthorized recipient, or risk $\geq 0.50$ $\rightarrow$ safe-view rewrite \\
Lease downgrade threshold & risk $\geq 0.30$ $\rightarrow$ lease downgrade \\
Label-free removal & ignores attack-applied flags, attack identifiers, attack modes, and evaluator- labels \\
\bottomrule
\end{tabular}
\caption{PAPC weights, thresholds, and override priorities used in the reported implementation.}
\label{tab:papc_params}
\end{table}

\begin{table}[!htbp]
\centering
\footnotesize
\setlength{\tabcolsep}{3pt}
\renewcommand{\arraystretch}{1.06}
\begin{tabular}{p{0.28\linewidth}p{0.42\linewidth}p{0.18\linewidth}}
\toprule
Raw item type & Safe abstraction / withholding behavior & Constructor \\
\midrule
Exact budget value & Replaced with a coarse statement that a budget constraint exists & Deterministic string rewrite \\
Internal timing/delay rationale & Replaced with coarse internal timing/delay abstraction when allowed & Deterministic string rewrite \\
Customer/project reference & Raw identifier withheld; safe text may state that a customer/project reference exists & Deterministic string rewrite \\
Credential-like marker & Raw marker withheld and represented as an internal credential-like marker withheld & Deterministic string rewrite \\
Unsafe instruction phrase & Replaced by a safe placeholder or routed to quarantine according to policy & Deterministic pattern rewrite/quarantine \\
\bottomrule
\end{tabular}
\caption{Safe-view construction. The evaluated implementation uses policy-driven deterministic rewriting. Exact raw-value checks validate that protected spans are removed.}
\label{tab:safeview_mapping}
\end{table}

\begin{table}[!htbp]
\centering
\footnotesize
\setlength{\tabcolsep}{3pt}
\renewcommand{\arraystretch}{1.06}
\begin{tabular}{p{0.30\linewidth}p{0.58\linewidth}}
\toprule
Mechanism & Evaluated behavior \\
\midrule
Quarantine store & Implemented for memory/workspace zones; quarantined raw content remains isolated from model-visible downstream channels \\
Downstream view after quarantine & Downstream agents receive a safe view, placeholder, or no emitted content depending on the decision path; raw content remains in quarantine \\
Propagation stop & Quarantine/block stops downstream propagation of the raw payload in the runtime event graph \\
Lease downgrade & Implemented as deterministic metadata/control signals such as downgrade or revoke, as benchmark control metadata \\
Lease scope in benchmark & Channel and abstraction restrictions narrow onward propagation; channel and abstraction restrictions implement the reported benchmark lease behavior \\
\bottomrule
\end{tabular}
\caption{Quarantine and lease semantics in the evaluated benchmark. The text avoids overstating lease narrowing as a full runtime capability system.}
\label{tab:quarantine_lease}
\end{table}

\begin{table}[!htbp]
\centering
\footnotesize
\setlength{\tabcolsep}{2.5pt}
\renewcommand{\arraystretch}{1.06}
\begin{tabular}{p{0.21\linewidth}p{0.51\linewidth}p{0.18\linewidth}}
\toprule
Metric & Implementation / definition & Scope \\
\midrule
Task success & Deterministic final-output rule/template check for required non-sensitive task content; reported separately from leakage & Final vendor-facing output \\
Unauthorized raw leakage & Exact raw-value string matching over event text, skipping quarantined/blocked events & Full runtime event sequence \\
External leakage & Raw leakage reaching external recipient, final output, vendor-send tool, or external message & External/final/tool channels \\
Cascade size/depth & Contaminated event-graph propagation size and longest contaminated causal path & Runtime event graph \\
Privilege reach & Maximum privilege touched by contaminated content, excluding quarantined/blocked events; privilege levels follow the runtime evaluator's 0--5 scale & Runtime event graph \\
Evaluator-label use & Count of label-free runs where defense metadata records use of evaluator-only attack labels & Defense decision metadata \\
\bottomrule
\end{tabular}
\caption{Metric implementation details. Leakage measurements are exact raw-value containment measurements.}
\label{tab:metric_impl}
\end{table}

\FloatBarrier

\section{Supplementary Method and Benchmark Tables}
\label{app:details}
These tables give compact implementation details and secondary benchmark summaries moved out of the main text to keep the submission focused.
\begin{table}[!htbp]
\centering
\footnotesize
\setlength{\tabcolsep}{3pt}
\renewcommand{\arraystretch}{1.08}
\begin{tabular}{p{0.26\linewidth}p{0.26\linewidth}p{0.36\linewidth}}
\toprule
Condition & Utility & Containment outcome \\
\midrule
No defense & Clean 0.373; attacked 0.333 & Exposed poisoned retrieval 0.467; manifestation 0.253. \\
PAPC & Clean 0.360; attacked 0.387 & Poison may be retrieved internally (0.440), but exposed poisoned retrieval and manifestation are 0. \\
Static keyword filter & Known-trigger comparator & Exposure and manifestation are 0 on this same-axis trigger; the result is used as same-axis trigger evidence. \\
\bottomrule
\end{tabular}
\caption{Adapted retrieval-memory exposure check. The table tests whether risky retrieved content becomes exposed downstream context or an unsafe action; it is an adapted retrieval-memory containment check.}
\label{tab:p0_agentpoison}
\end{table}

\begin{table}[!htbp]
\centering
\scriptsize
\setlength{\tabcolsep}{3.2pt}
\renewcommand{\arraystretch}{1.08}
\begin{tabular}{p{0.20\linewidth}p{0.21\linewidth}p{0.28\linewidth}p{0.21\linewidth}}
\toprule
Agent / channel & Principal / domain & Role in workflow & Raw-access policy \\
\midrule
Planner & Privileged internal coordinator & Plans the update, routes information, and coordinates finance and drafting through memory, messages, and summaries & May use authorized budget/delay/customer abstractions; may route  authorized abstractions to vendor paths \\
Finance agent & Privileged finance owner & Holds budget, internal rationale, customer reference, and credential-like protected items; supplies constraints for the update & Authorized raw access for finance-owned protected items, including credential-like marker \\
Document writer & Normal internal drafting principal & Produces internal draft/update text from planner and finance context using workspace and final-draft buffers & May receive internal drafting context; vendor-facing output must contain  allowed abstractions \\
External vendor endpoint & Low-trust external recipient/channel & Receives vendor-facing update through external message, vendor-send tool, or final output & Vendor-safe abstractions are allowed \\
\bottomrule
\end{tabular}
\caption{Agents, principals, and raw-access restrictions in the multi-agent workflow. Vendor-facing delivery is treated as an external channel even when represented by an agent object.}
\label{tab:agent_roles}
\end{table}

\begin{table}[!htbp]
\centering
\footnotesize
\setlength{\tabcolsep}{3pt}
\renewcommand{\arraystretch}{1.07}
\begin{tabular}{p{0.17\linewidth}p{0.33\linewidth}p{0.21\linewidth}p{0.19\linewidth}}
\toprule
Topology & Communication structure & Shared object & Propagation pressure \\
\midrule
chain\_4 & Planner $\rightarrow$ finance $\rightarrow$ document writer $\rightarrow$ external vendor & No global blackboard; next-hop messages and summaries dominate & Narrow but deeper handoff path \\
star\_4 & Planner hub connects finance, document writer, and external vendor & Coordinator state and hub summaries & Aggregation risk at central planner \\
blackboard\_4 & All-to-all interactions plus shared workspace/blackboard & Shared workspace artifact readable by multiple agents & Highest fanout and shared-state pressure \\
\bottomrule
\end{tabular}
\caption{Topology definitions. All topologies use the same task, protected items, attacks, defenses, and seeds; graph structure and shared-state fanout are the controlled variables.}
\label{tab:topologies}
\end{table}

\begin{table}[!htbp]
\centering
\footnotesize
\setlength{\tabcolsep}{3pt}
\renewcommand{\arraystretch}{1.08}
\begin{tabular}{p{0.12\linewidth}p{0.25\linewidth}p{0.25\linewidth}p{0.27\linewidth}}
\toprule
RQ & Evidence source & Setting & What it tests \\
\midrule
RQ1 & Adapted retrieval-memory comparator & P0 AgentPoison-style containment check & Whether risky retrieved content becomes exposed context or unsafe action before final output \\
RQ1--RQ2 & Deterministic propagation suite & Chain, star, and blackboard topologies & Intermediate propagation and topology-dependent cascade behavior \\
RQ3 & MiniMax final-writer evaluation & 252 runs: 3 topologies, 7 attack settings, 4 defenses, 3 seeds & Whether mediated context supports final writing while reducing exact raw exposure \\
RQ4 & Label-free paraphrase validation & 540 deterministic runs and 72 targeted MiniMax runs & Whether measured containment uses runtime-observable signals with evaluator labels removed \\
RQ4 & No-semantic-pattern ablation & Deterministic subset & Partial role of semantic sensitive-detail features \\
\bottomrule
\end{tabular}
\caption{Evidence sources organized by research question. The table clarifies the interpretation of each evidence slice.}
\label{tab:experiment_matrix}
\end{table}

\begin{table}[!htbp]
\centering
\footnotesize
\setlength{\tabcolsep}{3pt}
\renewcommand{\arraystretch}{1.08}
\begin{tabular}{p{0.22\linewidth}p{0.36\linewidth}p{0.30\linewidth}}
\toprule
Policy field & Meaning & Example \\
\midrule
Owner & Principal that owns the protected item & Finance team \\
Raw detector & Exact value or typed detector for protected content & Budget value, customer ID \\
Authorized raw readers & Principals allowed to receive the raw value & Planner and finance agents \\
Allowed abstraction & Safe downstream description & ``budget constraint exists'' \\
Forbidden channels & Destinations that may not carry raw content & Vendor output, shared blackboard \\
Sensitivity & Cost weight used for exposure accounting & High for credentials \\
\bottomrule
\end{tabular}
\caption{Policy interface. PAPC distinguishes forbidden raw disclosure from policy-allowed abstraction.}
\label{tab:policy_schema}
\end{table}

\begin{table}[!htbp]
\centering
\footnotesize
\setlength{\tabcolsep}{3pt}
\renewcommand{\arraystretch}{1.08}
\begin{tabular}{p{0.44\linewidth}p{0.20\linewidth}p{0.24\linewidth}}
\toprule
Runtime condition & Action & Effect \\
\midrule
Unauthorized raw content targets final, external, or tool channel & Block or safe view & Prevent direct exposure \\
Unauthorized raw content targets high-fanout shared state & Quarantine raw; emit safe view if valid & Stop cascade through shared object \\
Low-trust provenance enters high-privilege context & Lease downgrade & Narrow onward propagation rights \\
No raw violation and low propagation risk & Allow & Preserve useful task progress \\
Safe view cannot be validated & Quarantine & Fail closed for raw content \\
\bottomrule
\end{tabular}
\caption{Decision rules used by PAPC. Rules are fixed across reported experiments.}
\label{tab:decision_rules}
\end{table}

\begin{table}[!htbp]
\centering
\footnotesize
\setlength{\tabcolsep}{3pt}
\renewcommand{\arraystretch}{1.08}
\begin{tabular}{p{0.28\linewidth}p{0.36\linewidth}p{0.24\linewidth}}
\toprule
Channel & Example risk & Evaluation slice \\
\midrule
Retrieval memory & Poisoned memory becomes model-visible context & P0 comparator \\
Summary handoff & Private details copied into an agent summary & Summary poisoning \\
Workspace artifact & Shared document fans out sensitive content & Workspace poisoning \\
Inter-agent message & Low-trust message steers internal behavior & Comm. hijack \\
Paraphrased request & Dangerous request avoids known trigger phrases & Label-free validation \\
\bottomrule
\end{tabular}
\caption{Threat channels covered by the multi-agent runtime.}
\label{tab:threat_model}
\end{table}

\begin{table}[!htbp]
\centering
\footnotesize
\setlength{\tabcolsep}{5pt}
\renewcommand{\arraystretch}{1.08}
\begin{tabular}{lccc}
\toprule
Seed & Success & Raw leak & Ext. leak \\
\midrule
1 & 1.0 & 0.0 & 0.0 \\
2 & 1.0 & 0.0 & 0.0 \\
3 & 1.0 & 0.0 & 0.0 \\
All & 63/63 clean & 0.0 & 0.0 \\
\bottomrule
\end{tabular}
\caption{Seed stability in the MiniMax final-writer evaluation. PAPC remains clean across all evaluated PAPC runs for seeds 1--3.}
\label{tab:seed_stability}
\end{table}

\FloatBarrier

\section{Additional Evaluation Boundary and Safe-Trace Summaries}
\label{app:evidence}
The following tables summarize evaluation boundaries and redacted safe-trace cases.
\begin{table}[!htbp]
\centering
\footnotesize
\setlength{\tabcolsep}{4pt}
\renewcommand{\arraystretch}{1.06}
\begin{tabular}{p{0.20\linewidth}p{0.16\linewidth}p{0.20\linewidth}p{0.34\linewidth}}
\toprule
Setting & Scale & Outcome & Result \\
\midrule
Deterministic label-free & 540/540; no calls & Success 1.0; raw 0.0; external 0.0 & Zero evaluator-label use on evaluated paraphrase-validation matrices. \\
Targeted MiniMax label-free & 72/72; MiniMax calls & Success 1.0; raw 0.0; external 0.0 & Final-writer validation with zero evaluator-label use in the evaluated variants. \\
Det. vs. no defense & 90 comparisons & Raw 81 improved / 9 tied; external 81 improved / 9 tied & No leakage underperformance in matched groups. \\
Det. vs. static ACL & 90 comparisons & Raw 66 improved / 24 tied; external 48 improved / 42 tied & ACL policy gaps remain. \\
Det. vs. prompt filter & 90 comparisons & Raw 54 improved / 36 tied; external 54 improved / 36 tied & Paraphrase failures are recorded for prompt filtering. \\
MiniMax comparisons & 18 comparisons each & Raw: 15/15/16 improved; external: 10/9/10 improved & Final-writer comparisons against no defense, static ACL, and prompt filter. \\
No-pattern ablation & deterministic subset & Success tied; raw higher; external tied & Semantic request features reduce raw exposure; policy/fanout/safe-view still contribute. \\
\bottomrule
\end{tabular}
\caption{Label-free paraphrase validation and partial ablation. The label-free variant removes evaluator-only attack annotations from the guard input; the result reports measured containment for the evaluated paraphrase variants.}
\label{tab:nonoracle}
\end{table}

\begin{table}[!htbp]
\centering
\footnotesize
\setlength{\tabcolsep}{4pt}
\renewcommand{\arraystretch}{1.08}
\begin{tabular}{p{0.18\linewidth}p{0.31\linewidth}p{0.31\linewidth}p{0.11\linewidth}}
\toprule
Evaluation slice & Supports & Boundary & Scale \\
\midrule
P0 adapted comparator & Retrieval-memory containment on a saved adapted full-ReAct axis. & Retrieval-memory containment transfer beyond the adapted full-ReAct axis.  & summaries \\
Deterministic MAS & Propagation mechanism, topology effects, and workflow indirect-attack comparisons. & Real-provider behavior. & 252 runs \\
MiniMax final-writer evaluation & MiniMax final-writer workflow evaluation, PAPC clean subset, and improvement/tie comparisons. & Additional provider and real computer-use validation. & 252 runs \\
Label-free validation & Reserved paraphrase variant evidence without evaluator-only attack annotations. & Additional paraphrase and adaptive-adversary validation. & 540 det. + 72 MiniMax \\
Mechanism ablation & Semantic patterns contribute to raw-leakage containment; policy/fanout/safe-view still matter. & Full component-isolation study. & partial \\
\bottomrule
\end{tabular}
\caption{Evaluation scope. The table links each experimental slice to its supported interpretation and evaluation boundary.}
\label{tab:evidence_boundaries}
\end{table}

\begin{table}[!htbp]
\centering
\footnotesize
\setlength{\tabcolsep}{4pt}
\renewcommand{\arraystretch}{1.08}
\begin{tabular}{p{0.20\linewidth}p{0.24\linewidth}p{0.36\linewidth}p{0.12\linewidth}}
\toprule
Illustration & Setting & Safe-trace observation & Role \\
\midrule
No-defense workspace leak & Blackboard; indirect workspace; seed 1; no defense & Shared workspace propagation reaches an external-facing path; task succeeds, but raw leak 16, external leak 1, cascade 7, privilege reach 5. & Failure case \\
Prompt-filter paraphrase failure & Blackboard; paraphrased workspace; seed 1; prompt filter & Paraphrased shared-state request bypasses surface-form filtering; raw leak 16 and external leak 1 in the targeted safe trace. & Baseline note \\
PAPC safe-view/quarantine & Blackboard; indirect workspace; seed 1; PAPC & A risky workspace write is quarantined or rewritten; task success is preserved and raw/external leakage and privilege reach are 0. & Prevention pair \\
Label-free reserved paraphrase success & Blackboard; paraphrased workspace; seed 1; label-free PAPC & Quarantine occurs with evaluator-label use false; task success true, raw/external leakage 0, and evaluator-label use 0. & Validity note \\
\bottomrule
\end{tabular}
\caption{Redacted qualitative examples derived from safe traces. They illustrate evaluated MiniMax final-writer workflow-benchmark behavior; they are redacted safe-trace summaries for interpretation.}
\label{tab:case_studies}
\end{table}

\FloatBarrier

\end{document}